\documentclass[12pt]{article}
\usepackage{lineno}
\usepackage{amsthm}
\usepackage{here}
\usepackage{natbib}
\usepackage{lscape}
\usepackage{longtable}
\usepackage{bm,graphicx,graphics,url,psfrag}
\usepackage{lipsum} 

\usepackage{amsthm,amsmath,amssymb,amsfonts,mathrsfs,mathtools}
\usepackage{breakcites}

\usepackage{adjustbox}
\usepackage{rotating}
\usepackage{subfigure}

\usepackage[paperwidth=26cm]{geometry}

\usepackage{multirow}

\usepackage{listings}
\usepackage{color}

\definecolor{codegreen}{rgb}{0,0.6,0}
\definecolor{codegray}{rgb}{0.5,0.5,0.5}
\definecolor{codepurple}{rgb}{0.58,0,0.82}
\definecolor{backcolour}{rgb}{0.95,0.95,0.92}

\lstdefinestyle{mystyle}{
    backgroundcolor=\color{backcolour},
    commentstyle=\color{codegreen},
    keywordstyle=\color{magenta},
    numberstyle=\tiny\color{codegray},
    stringstyle=\color{codepurple},
    basicstyle=\footnotesize,
    breakatwhitespace=false,
    breaklines=true,
    captionpos=b,
    keepspaces=true,
    numbers=left,
    numbersep=5pt,
    showspaces=false,
    showstringspaces=false,
    showtabs=true,
    tabsize=2
}

\title{A class of asymmetric regression models for left-censored data}

 \author{\normalsize
  \textbf{Helton Saulo}$^{1}$, \textbf{Jeremias Le\~ao}$^{2}$\thanks{Corresponding
author: Jeremias Le\~ao. Email: leaojeremiass@gmail.com}\,\,, \textbf{Juvencio Nobre}$^{3}$, \textbf{N. Balakrishnan}$^{4}$\\
 { $^{1}$Department of Statistics, Universidade de Bras\'ilia, Bras\'ilia, DF, Brazil}\\
  {$^{2}$Department of Statistics, Universidade Federal do Amazonas, Amazonas, MA, Brazil}\\
 {$^{3}$Department of Statistics, Universidade Federal do Cear\'a, Fortaleza, CE, Brazil}\\
 {$^{4}$Department of Mathematics and Statistics, McMaster University, Hamilton, ON, Canada}
 }

\begin{document}
\maketitle

\begin{abstract}
A common assumption regarding the standard tobit model is the normality of the error distribution. 
However, asymmetry and bimodality may be present and alternative tobit models must be used. In this 
paper, we propose a tobit model based on the class of log-symmetric distributions, which includes as
special cases heavy and light tailed distributions and bimodal distributions. We implement a likelihood-based 
approach for parameter estimation and derive a type of residual. We then discuss the problem of performing 
testing inference in the proposed class by using the likelihood ratio and gradient statistics, which 
are particularly convenient for tobit models, as they do not require the information matrix. A thorough Monte Carlo study
is presented to evaluate the performance of the maximum likelihood estimators and the likelihood ratio and gradient tests. Finally, 
we illustrate the proposed methodology by using a real-world data set.

\begin{keywords}
Log-symmetric distributions; Likelihood ratio test; Gradient test; \texttt{R} software; Tobit models.
\end{keywords}
\end{abstract}

\section{Introduction}

After its introduction by \cite{t:58}, the tobit model has been used extensively in several areas including economics, environmental sciences, 
engineering, biology, medicine and sociology; see, for example, \cite{bpl:08}, \cite{lbpg:07}, \cite{vpl:11}, \cite{amemiya:84}, \cite{helsel:11}, 
\cite{thordisg:10} and \cite{martinezetal:13b,martinezetal:13a}. The tobit model is used to described censored responses and 
had its motivation based on a study to analyze the relationship between household expenditure on a durable good and household incomes. In this study, 
\cite{t:58} faced the existence of many cases where the expenditure was zero, which violated the linearity assumption of common regression approaches. 
\cite{t:58} introduced a regression model whose response was censored at a prefixed limiting value; see \cite{amemiya:84}.

A strong assumption of the tobit model is that the error term is normality distributed, but it is not always the case in many applications; see, for example, 
\cite{bggl:10,bgls:17}. The normality assumption may not be appropriate to describe the behavior of strictly positive data, as well as bimodal and/or 
light- and heavy-tailed data. The use of flexible distributions is very important as often real-world data are better modeled by non-normal 
distributions, especially in the aspect related to the robustness of the results. In the context of censored responses, some authors have emphasized the importance to use more flexible distributions;     
see, for example, \cite{acgm:12}, \cite{martinezetal:13b,martinezetal:13a}, \cite{gblc:15}, \cite{massuiaetal:15} 
and \cite{bggl:10,bgls:17}.

The log-symmetric distribution class was investigated by \cite{j:08} and arises when a {random variable (RV)} has the same distribution as its reciprocal 
or when the distribution of the logged {RV} is symmetrical. This class is very useful for modeling strictly positive, asymmetric, bimodal and 
light- and heavy-tailed data. The class of log-symmetric distributions is a generalization of the log-normal distribution, which provides more 
flexible alternatives; see, for example, \cite{vanegasp:16a}. \cite{vanegasp:15} proposed a semiparametric regression
model allowing both median and skewness to be modeled, \cite{vanegasp:16a} discussed some statistical properties of the log-symmetric class of distributions, 
\cite{vanegasp:16b} proposed an extension of the log-symmetric regression models used by \cite{vanegasp:15} considering  an arbitrary number of
non-parametric additive components to describe the median and skewness and \cite{medeirosferrari:16} 
considered the issue of testing hypothesis in symmetric 
and log-symmetric linear regression models.

A prominent and recent procedure for hypothesis testing in parametric models is the {gradient (GR)} test, 
which was proposed by \cite{terrell:02}. 
This procedure is simple to compute and only involves the score vector and the maximum likelihood (ML) 
estimates of the parameter vector under 
the unrestricted and restricted models. Similarly to the generalized likelihood ratio (LR) statistic \citep{w:38}, the 
{GR} statistic is also attractive for censored 
samples, as is the case of tobit models, since no computation of the information matrix 
(neither observed nor expected) is required; see, for example, \cite{lf:11d}. 

In this context, the primary objective of this paper is to propose a class of tobit models 
based on the log-symmetric distribution. The secondary objectives are:  (i) to obtain the ML estimators of the model parameters; 
(ii) to deal with the issue of performing hypothesis testing concerning the parameters of the proposed model. The {LR} 
and {GR} tests are used for the hypothesis testing purpose; (iii) to carry out {Monte Carlo (MC)} simulations to evaluate 
the performances of ML estimators and the {LR} and {GR} tests; 
and (iv) to conduct a real world data application of the proposed methodology.

The rest of the paper proceeds as follows. In Section \ref{sec:2}, 
we describe briefly the class of log-symmetric distributions and 
some properties. In Section \ref{sec:3}, we formulate the tobit model based 
on the log-symmetric class, provide estimation, inference 
and residual analysis based on the ML method. In Section \ref{sec:4}, we carry out the mentioned MC
simulations and an empirical application with real-world data is done in Section \ref{sec:5}. Finally, 
in Section \ref{sec:6}, we discuss conclusions and future research on the topic of this work.

\section{Log-symmetric distributions}\label{sec:2}

Consider a continuous and symmetric RV $Y$ having a symmetric distribution with location parameter 
$\mu\in\mathbb{R}$, dispersion parameter $\phi>0$, density generator $g(\cdot)$ and probability density function (PDF)
\begin{equation*}\label{eqls:sympdf}
 f_{Y}(y;\mu,\phi,g)=\frac{1}{\phi}g\left(\frac{(y-\mu)^2}{\phi^2} \right), \quad y\in\mathbb{R},
\end{equation*}
with $g(u)>0$ for $u>0$ and $\int_{0}^{\infty}u^{-1/2}g(u)\partial{u}=1$; 
see \cite{fkn:90}. In this case, the notation $Y\sim\textrm{S}(\mu,\phi^2,g)$ is used. 
The class of log-symmetric distributions arises when we set $T=\exp(Y)$, that is, we obtain a continuous 
and positive {RV} $T$ 
such that the distribution of its logarithm belongs to the symmetric 
family. The PDF of $T$ is written as
\begin{equation*}\label{eqls:logsympdf}
 f_{T}(t;\eta,\phi,g)=\frac{1}{{\phi}t}g(\widetilde{t}^{\,2}), \quad t>0,
\end{equation*}
where $\widetilde{t}=\log\left( [t/\eta]^{{1}/{{\phi}}} \right)$ and $\eta=\exp(\mu)>0$ 
is a scale parameter. We write $T\sim\textrm{LS}(\eta,\phi^2,g)$. The density generator $g$ may be associated 
with an extra parameter $\xi$ (or an extra parameter vector $\bm\xi$). The 
cumulative distribution function (CDF) of $T$ is given by
\begin{equation*}\label{eqls:logsymcdf}
 F_{T}(t;\eta,\phi,g)=F_{Z}(\widetilde{t};\mu,\phi,g),
\end{equation*}
where $F_{Z}(\cdot)$ is CDF of $Z=(Y-\mu)/{\phi}\sim\textrm{S}(0,1,g)$. 

Note that the density generator 
$g$ leads to different log-symmetric distributions. Some members of log-symmetric distributions 
are the log-normal \citep{cs:88,jkb:94}, log-logistic \citep{mo:07}, 
log-Laplace \citep{jkb:95}, 
log-Cauchy \citep{mo:07}, log-power-exponential \citep{vanegasp:16a},
log-Student-$t$ \citep{vanegasp:16a}, log-power-exponential \citep{vanegasp:16a}, log-slash \citep{vanegasp:16a}, 
harmonic law \citep{p:08}, Birnbaum-Saunders \citep{bs:69a,rn:91},
generalized Birnbaum-Saunders \citep{dl:05}, and F \citep{jkb:95} distributions; see Table~\ref{tab:1}.

Let $T\sim\textrm{LS}(\eta,\phi^2,g)$, then we have the properties: (P1) $cT\sim\textrm{LS}(c\eta,\phi^2,g)$, with $c>{0}$; (P2) $T^{c}\sim\textrm{LS}(\eta^{c},c^{2}\phi^2,g)$, with $c\neq{0}$; 
and (P3) the median of the distribution of $T$ is $\eta$. The properties (P1) and (P2) say that the log-symmetric distribution holds 
the proportionality and reciprocation properties, respectively. Moreover, (P2) is useful to propose modified moment estimators; 
see \cite{nkb:03} for the Birnbaum-Saunders case. Finally, (P3) can be used to specify a dynamic point process model in terms of 
the conditional median; see \cite{saulolla:16}.

\begin{table}[htbp] 
\centering
\small
\caption{\small{Density generator $g(u)$ for some log-symmetric distributions.}}\label{tab:1}
\begin{tabular}{lcccccccccc}
\hline Distribution                          &&             $g(u)$                                                              \\ 
\hline
Log-normal($\eta,\phi$)                    && $\propto\exp\left( -\frac{1}{2}u\right)$                                       \\[1.5ex]
Log-Student-$t$($\eta,\phi,\xi$)           && $\propto\left(1+\frac{u}{\xi} \right)^{-\frac{\xi+1}{2}}$, $\xi>0$               \\[1.5ex]
Log-power-exponential($\eta,\phi,\xi$)     && $\propto\exp\left( -\frac{1}{2}u^{\frac{1}{1+\xi}}\right)$, $-1<{\xi}\leq{1}$   \\[1.5ex]
Birnbaum-Saunders($\eta,\phi=4,\xi$)         && $\propto\cosh(u^{1/2})\exp\left(-\frac{2}{\xi^2}\sinh^2(u^{1/2}) \right) $, $\xi>0$ \\[1.5ex]
Birnbaum-Saunders-$t$($\eta,\phi=4,{\bm\xi}=(\xi_{1},\xi_{2})^{\top}$) && $\propto\cosh(u^{1/2})\left(\xi_{2}\xi_{1}^2+4\sinh^2(u^{1/2})\right)^{-\frac{\xi_{2}+1}{2}} $, $\xi_{1},\xi_{2}>0$ \\[1.5ex]
\hline
\end{tabular}
\end{table}

\section{The tobit-log-symmetric model}\label{sec:3}

Consider a censored response variable to the left $Y_{i}$ for the case $i$, 
which is observable for values greater than $\gamma$ and censored 
for values smaller than or equal to $\gamma$. Then, in the tobit formulation 
\begin{equation}\label{eq:tobitnormal}
Y_{i} = \begin{cases}
\gamma, & \quad Y_{i}^{\ast} \;  \leq \; \gamma, \quad i = 1, \ldots,m;\\
\bm{x}_{i}^{\top}{\bm\beta} + \varepsilon_{i}, & \quad Y_{i}^{\ast} \; > \; \gamma, \quad i = m+1, \ldots,n,
\end{cases}
\end{equation}
where $Y_{i}^{\ast} =\bm{x}_{i}^{\top}{\bm\beta} + \varepsilon_{i}$, $m$ is the number of cases censored to the left, $n$ 
is the total number of cases, $\bm{x}_{i}=(x_{i1},\ldots,x_{ip})^{\top}$ is an $n\times{1}$ vector of covariates fixed and known, 
${\bm\beta}=(\beta_{1},\dots,\beta_{p})^{\top}$ is a $p\times{1}$ vector of regression coefficients, and $\{\varepsilon_{i}\}$ are independent identically 
distributed (IID) {RVs}.  The tobit-normal (tobit-NO) model is obtained from \eqref{eq:tobitnormal} when $\varepsilon_{i}$ follows a normal 
distribution with mean zero and variance $\varsigma^2$, that is, $\varepsilon_{i} {\stackrel{\textrm{\tiny IID}}{\sim}} \textrm{N}(0, \varsigma^2)$.


Consider the log-symmetric regression model \citep{vanegasp:15}
\begin{equation}\label{eq:logsym}
T_{i}=\eta_{i}\,\epsilon_{i}^{{\phi_{i}}}, \quad i=1,\ldots,n,
\end{equation}
where $\eta_{i}$ and $\phi_{i}$ are median and skewness of the $T_{i}$ distribution, respectively, and 
$\{\epsilon_{i}\}$ are standard log-symmetric distributed IID {RVs} denoted by $\epsilon_{i} {\stackrel{\textrm{\tiny IID}}{\sim}} \textrm{LS}(1, 1, g)$. 
Then, $T_i {\stackrel{\textrm{\tiny IND}}{\sim}} \textrm{LS}(\eta_{i},\phi_{i}^2,g)$. The structures for $\eta_{i}$ and $\phi_{i}$ are written as
\begin{eqnarray}\label{eq_sctr1}
\eta_{i}&=&\exp(\bm{x}_{i}^{\top}{\bm\beta}),\quad i=1,\ldots,n, \nonumber\\
\log(\phi_{i})&=& {\bm w}_{i}^{\top}{\bm\zeta}, \quad i=1,\ldots,n,\nonumber \label{eq_scr2}
\end{eqnarray}
where $\bm{x}_{i}$ and ${\bm\beta}$ are as in \eqref{eq:tobitnormal}, ${\bm w}_{i}=(w_{ik},\ldots,w_{ik})$ is an $n\times{1}$ 
vector of covariates for $\phi_{i}$ and ${\bm\zeta}=(\zeta_{1},\ldots,\zeta_{k})^{\top}$ is a $p\times{1}$ parameter 
vector. For simplicity reasons, hereafter it is assumed that $\phi_{i}=\phi$, for ${i}=1,\ldots,n$.

By applying logarithm  in Equation \eqref{eq:logsym}, we obtain
\begin{equation}\label{eq:logsymtobit}
 \underbrace{\log(T_{i})}_{Y_{i}}= \underbrace{\log(\eta_{i})}_{\mu_{i}}+{\phi}\underbrace{\log(\epsilon_{i})}_{\varepsilon_{i}}, \quad i = 1, \ldots, n,
\end{equation}
where $\varepsilon_i$ is standard symmetric distributed, $\varepsilon_i {\stackrel{\textrm{\tiny IID}}{\sim}} \textrm{S}(0,1,g)$, 
and $Y_i {\stackrel{\textrm{\tiny IND}}{\sim}} \textrm{S}(\mu_{i},\phi^2,g)$. Then, based on Equations \eqref{eq:tobitnormal} and \eqref{eq:logsymtobit},  
we propose a tobit model based on the log-symmetric distribution, denoted by tobit-LS, as
\begin{eqnarray}\label{eq:tobit-logsym}
Y_{i}=
\begin{cases}
\gamma, & \quad Y_{i}^{\ast}  \leq  \gamma, \quad i=1,\ldots,m;\\
\bm{x}_{i}^{\top}{\bm\beta} + \varepsilon_{i}, & \quad  Y_{i}^{\ast}  >  \gamma, \quad i=m+1,\ldots,n;
\end{cases}
\end{eqnarray}
where $Y_{i}^{\ast}=\log(T_{i}^{\ast})=\bm{x}_{i}^{\top}{\bm\beta} + \varepsilon_{i}$, 
${\bm \beta}$ and $\bm{x}_{i}$ are as in \eqref{eq:tobitnormal}, and $\varepsilon_{i}$ is as in \eqref{eq:logsymtobit}.

Consider a sample of size $n$, ${\bm Y}=(Y_1,\ldots,Y_m,Y_{m+1},\ldots,Y_n)^{\top}$ say, from a tobit-LS model that contains $m$ 
left-censored data, that is, the values of $Y$ less than a threshold point $\gamma$, and $n-m$ complete or uncensored data, namely, 
values of $Y$ greater than $\gamma$. Then, the corresponding likelihood function for   
 ${\bm\theta} = ({\bm \beta}^{\top},\phi)^{\top}$
\begin{equation*}\label{eq:loglik1}
L({\bm\theta})= \prod_{i=1}^{m}  F_{Y}(\zeta^{\textrm{c}}_{i};\mu_{i},\phi,g) \prod_{i=m+1}^{n} \frac{1}{\phi}g\left( \zeta_{i}^2 \right),
\end{equation*}
where $F_{Y}$ is the CDF of the symmetric distribution and 
\begin{equation}\label{eq:loglik2}
 \zeta^{\textrm{c}}_{i}=\left(\frac{\gamma - {\bm x_{i}^{\top}{\bm \beta}}}{\phi}\right)  \quad \text{and} \quad \zeta_{i}=\left(\frac{y_{i} - {\bm x_{i}^{\top}{\bm \beta}}}{\phi}\right).
\end{equation}

By taking the logarithm of \eqref{eq:loglik2}, we obtain the log-likelihood function for ${\bm\theta} = ({\bm \beta}^{\top},\phi)^{\top}$, which is given by
\begin{equation}\label{eq:loglik3}
\ell({\bm\theta}) = \sum_{i} \ell_i({\bm\theta}),
\end{equation}
where
$$
\ell_i({\bm\theta}) =
\begin{cases}
\log(F_{Y}(\zeta^{\textrm{c}}_{i};\mu_{i},\phi,g)), & i=1,\ldots,m;\\[0.25cm]
-\log(\phi)+\log(g\left( \zeta_{i}^2 \right)), &  i=m+1,\ldots,n.
\end{cases}
$$

The score vector for ${\bm \beta}$ and $\phi$ is given by 
\begin{equation}\label{eq:score}
\dot{\bm \ell}({\bm\theta})= \dfrac{\partial{\ell({\bm\theta})}}{\partial{{\bm\theta}}} = \sum_{i=1}^{n} \dot{\bm \ell}_i({\bm\theta}), \quad
\text{where}\,\, \dot{\bm \ell}_i({\bm\theta}) = (\dot{\bm \ell}_{i{\bm \beta}}^\top({\bm\theta}),\dot{\ell}_{i{\phi}}({\bm\theta}))^\top 
\end{equation}
with
$$
\dot{\bm\ell}_{i{\bm\beta}}({\bm\theta}) =
\begin{cases}
 -\frac{1}{\phi} \Omega_{i} {\bm x_{i}},  & i=1,\ldots,m;\\[0.25cm]
  -\frac{2}{\phi}W_{i}\zeta_{i} {\bm x_{i}} , & i=m+1,\ldots,n;
\end{cases}
$$

%
$$
\dot{\ell}_{i{\phi}}({\bm\theta}) =
\begin{cases}
 -\frac{1}{\phi}\Omega_{i}\zeta^{\textrm{c}}_{i}, & i=1,\ldots,m;\\[0.25cm]
 -\frac{1}{\phi}-\frac{2}{\phi}W_{i}\zeta_{i}^2, & i=m+1,\ldots,n;
\end{cases}
$$
with
$\Omega_i=\frac{{\rm d}F_{Y}(u)/{\rm d}u|_{u=\zeta^{\textrm{c}}_{i}}}{F_{Y}(\zeta^{\textrm{c}}_{i})} $ and
$W_{i}=\frac{{\rm d}g(u)/{\rm d}u|_{u=\zeta_{i}^2}}{g(\zeta_{i}^{2})} $. To obtain the ML estimate of ${\bm\theta}$ 
it is necessary to maximize the expression defined in 
\eqref{eq:loglik3} by equating the score vector $\dot{\bm \ell}({\bm\theta})$ to zero, providing the 
likelihood equations. They are solved using the Broyden-Fletcher-Goldfarb-Shanno (BFGS) quasi-Newton method; 
see \citet[][p.\,199]{mjm:00}. The corresponding standard errors (SEs) can be approximated by computing the 
square roots of the diagonal elements of the inverse of the observed Fisher information matrix \citep{eh:78}, 
which is obtained as  ${\cal J}({\bm \theta})= -\ddot{\bm \ell}({\bm\theta})$, where $\ddot{\bm \ell}({\bm\theta})$ 
denotes the Hessian matrix, that is,
\begin{equation*}\label{eq:hessian}
\ddot{\bm \ell}({\bm\theta})= \dfrac{\partial^{2}{\ell({\bm\theta})}}{\partial{{\bm\theta}}\partial{{\bm\theta}^{\top}}} = \sum_{i=1}^{n} \ddot{\bm \ell}_i({\bm\theta}), \quad
\text{where}\,\, 
\ddot{\bm \ell}_i({\bm\theta}) =
\left[\begin{array}{cc}
 \ddot{\ell}_{i{\bm\beta}{\bm\beta}}({\bm\theta}) &   \ddot{\ell}_{i{\bm\beta}\phi}({\bm\theta})\\
  \ddot{\ell}_{i\phi{\bm\beta}}({\bm\theta}) &   \ddot{\ell}_{i\phi\phi}({\bm\theta})
\end{array}\right],
\end{equation*}
with
$$
\ddot{\bm\ell}_{i{\bm\beta}{\bm\beta}}({\bm\theta}) =
\begin{cases}
 -\frac{1}{\phi} \Omega_{i}'{\bm x_{i}} ,  & i=1,\ldots,m;\\[0.25cm]
  -\frac{2}{\phi}\left[ -W_{i}\frac{{\bm x_{i}}}{\phi}  + W_{i}'\zeta_{i} \right] {\bm x_{i}} , & i=m+1,\ldots,n;
\end{cases}
$$
$$
\ddot{\bm\ell}_{i{\bm\beta}\phi}({\bm\theta}) = \ddot{\bm\ell}_{i\phi{\bm\beta}}({\bm\theta})=
\begin{cases}
 - \left[ \frac{1}{\phi}\Omega_{i}'-\frac{1}{\phi^{2}}\Omega_{i} \right] {\bm x_{i}},  & i=1,\ldots,m;\\[0.25cm]
  -\frac{2\zeta_{i}}{\phi}\left\{ \left[ -\frac{W_{i}}{\phi} + W_{i}'\right]  -\frac{1}{\phi}W_{i}    \right\} {\bm x_{i}} , & i=m+1,\ldots,n;
\end{cases}
$$
$$
\dot{\ell}_{i{\phi}{\phi}}({\bm\theta}) =
\begin{cases}
 \frac{1}{\phi^{2}} [2 \Omega_{i}-\phi \Omega_{i}' ] \zeta^{\textrm{c}}_{i}, & i=1,\ldots,m;\\[0.25cm]
 \frac{1}{\phi^{2}}+\frac{2}{\phi^{2}}[3W_{i}-\phi W_{i}']\zeta_{i}^2, & i=m+1,\ldots,n.
\end{cases}
$$

\subsection{Statistical tests}

We here consider the LR and GR statistical tests for the tobit-log-symmetric regression model. We choose these tests because they do not require 
the information matrix, a convenient characteristic for tobit models. 
Let ${\bm \theta}$ be a $p$-vector of parameters that index a tobit-log-symmetric model. Consider that our
interest lies in test the hypothesis ${\cal{H}}_{0}: {\bm \theta}_{1}={\bm \theta}^{(0)}_{1}$
against ${\cal{H}}_{1}: {\bm \theta}_{1}\neq{\bm \theta}^{(0)}_{1}$,
where ${\bm \theta}=({\bm \theta}^{\top}_{1},{\bm \theta}^{\top}_{2})^{\top}$, ${\bm \theta}_{1}$ 
is an $r \times 1$ vector of parameters of interest 
and ${\bm \theta}_{2}$ is $(p-r) \times 1$ vector of 
nuisance parameters.

Two popular methods for testing these linear hypotheses 
are by using the {LR} and {GR} test statistics, which are given by
\begin{align*}
\Lambda_{LR} &= 2\{\ell(\widehat{\bm \theta})-\ell(\widetilde{\bm \theta})\},\nonumber\\
\Lambda_{GR}  &= \dot{\bm \ell}^{\top}(\widetilde{\bm \theta})(\widehat{\bm \theta}-\widetilde{\bm \theta}), 
\end{align*}
where $\ell(\cdot)$ is the log-likelihood function defined in \eqref{eq:loglik3} and $\widehat{\bm \theta}=(\widehat{\bm \theta}^{\top}_{1},\widehat{\bm \theta}^{\top}_{2})^{\top}$ and
$\widetilde{\bm \theta}=({\bm \theta}^{(0)\top}_{1},\widetilde{\bm \theta}^{\top}_{2})^{\top}$
are unrestricted and restricted ML estimators of ${\bm \theta}$, respectively. Moreover,
$\dot{\bm \ell}(\cdot)$ is the the score vector defined in \eqref{eq:score}. In regular cases, we have that under ${\cal{H}}_{0}$ and $n\rightarrow\infty$, both statistical tests converge in distribution
to $\chi^{2}_{r}$. Then, ${\cal{H}}_{0}$ is rejected at nominal 
level $\delta$ if the test statistic is larger than
$\chi^{2}_{1-\delta,r}$, the $1-\delta$ 
upper quantile of the $\chi^{2}_{r}$ distribution.


\subsection{Model checking }

Residuals analysis are frequently used to evaluate the validity of the assumptions of the model, presence of outliers and may also be employed as tools for model selection. In the context of 
regression models, usually Pearson and studentized residuals are often used. Nevertheless, in a tobit scenario, these two types of residuals, even under normality, are not inadequate; see, for example, \citet{bggl:10}. In the log-symmetric tobit case, we use the generalized Cox-Snell (GCS) residual given by
$$
r_{i}^{\textrm{GCS}}=-\log(\widehat{S}_{Y}(y_{i};\widehat{\mu}_{i},\widehat{\phi}^2,g))=-\log(1-\widehat{F}_{Y}(y_{i};\widehat{\mu}_{i},\widehat{\phi}^2,g)),  \quad i=1,\ldots,n,
$$
where $\widehat{S}_Y$ denotes survival function fitted to the data. The GCS residual is unit exponential, $\textrm{EXP}(1)$ in short, if the model is correctly specified 
whatever the specification of the model.

\section{Monte Carlo simulation studies}\label{sec:4}
Two {MC} simulation studies were carried out to evaluate 
the performances of the ML estimators and the statistical tests. We focus on 
three tobit-log-symmetric models: tobit-log-normal (tobit-LN), tobit-log-Student-$t$ (tobit-L$t$) 
and tobit-log-power-exponential (tobit-LPE). The 
\texttt{R} software was used to do all numerical calculations; see \cite{r:14}.

\subsection{ML estimators}
A {MC} simulation study was carried out to evaluate the performance of the ML estimators. The study 
considers simulated data generated from each one of the above-mentioned models according to
\begin{equation*}
Y_{i} =
\begin{cases}
   \gamma, &  Y_{i}^{\ast} \leq  \gamma,\,\, i=1,\ldots, m,\\
  Y_{i}^{\ast}= \beta_{0} + \beta_{1}x_{i}+\varepsilon_{i}, & Y_{i}^{\ast}> \gamma,\,\, i=m+1,\ldots,n,\\
\end{cases}
\end{equation*}
where $\varepsilon_{i}$ is as in \eqref{eq:tobit-logsym}, $x_{i}$ is a covariate 
obtained from a uniform distribution in the interval (0,1) 
and the true parameter values are taken as $\beta_{0}=0.2$ $\beta_{1}=0.5$. 
Moreover, the simulation scenario considers: sample size $n \in \{50, 100, 300, 500\}$, scale parameter 
$\phi \in \{1.00, 3.00, 5.00\}$, 
extra parameter $\xi_{1}=0.5$ (tobit-LPE), $\xi_{1}=4$ (tobit-L$t$), censoring proportion $\varrho=m/n \in \{0.20, 0.50\}$, 
with 5,000 {MC} replications for each sample size.

The ML estimation results for the considered tobit-log-symmetric models are presented in Tables \ref{table:MC1}--\ref{table:MC3}. 
The empirical bias and mean squared error (MSE) are reported. A look at the results in Tables \ref{table:MC1}--\ref{table:MC3} allows us to conclude that, 
for $\phi \in \{1.00, 3.00, 5.00\}$ and $\varrho \in \{0.20, 0.50\}$, as the sample size
increases, the empirical bias and MSE decrease, as expected. Moreover, we note that, as the value of the parameter 
$\phi$ increases, the performance of the estimator of this parameter, deteriorates. In general, 
the performances of the estimators decrease when the censoring proportion increases.

\begin{table}[!ht]
\small
\centering
 \renewcommand{\arraystretch}{0.8}
 \renewcommand{\tabcolsep}{0.2cm}
\caption{
Empirical bias and MSE (in parentheses) from simulated data for the indicated ML estimators of the tobit-LN 
model parameters, $n$ and $\varrho$.} \label{table:MC1}
\begin{adjustbox}{max width=\textwidth}
  \begin{tabular}{lrrrrrrrrrr}
	\hline\vspace{-0.1cm} \\
$n$  &&&  \multicolumn{3}{c}{$\varrho =0.20$} &&& \multicolumn{3}{c}{$\varrho=0.50$}  \\
   \cline{4-6} \cline{9-11} \vspace{-0.1cm} \\
& \multicolumn{1}{l}{$\phi$} && \multicolumn{1}{c}{$\widehat\phi$} & \multicolumn{1}{c}{$\widehat\beta_{0}$} & \multicolumn{1}{c}{$\widehat\beta_{1}$} &&&
\multicolumn{1}{c}{$\widehat\phi$} & \multicolumn{1}{c}{$\widehat\beta_{0}$} & \multicolumn{1}{c}{$\widehat\beta_{1}$} \\
  \hline \vspace{-0.1cm} \\
50    & 1.00  && $-$0.0099\,(0.0141) & $-$0.0147\,(0.0918) & 0.0086\,(0.2715)  &&& $-$0.0132\,(0.0249) & $-$0.0212\,(0.1218) & 0.0087\,(0.3263)\\
      & 3.00  && $-$0.0297\,(0.1269) & $-$0.0367\,(0.8167) & 0.0114\,(2.4409)  &&& $-$0.0394\,(0.2250) & $-$0.0589\,(1.0393) & 0.0169\,(2.9163)\\
      & 5.00  && $-$0.0491\,(0.3526) & $-$0.0589\,(2.2629) & 0.0144\,(6.7631)  &&& $-$0.0652\,(0.6263) & $-$0.0951\,(2.8684) & 0.0210\,(8.1019)\\
\\
100   & 1.00  && $-$0.0045\,(0.0068) & $-$0.0092\,(0.0441) & 0.0072\,(0.1308)  &&& $-$0.0084\,(0.0125) & $-$0.0094\,(0.0573) & 0.0060\,(0.1518)\\
      & 3.00  && $-$0.0138\,(0.0610) & $-$0.0239\,(0.3924) & 0.0148\,(1.1741)  &&& $-$0.0257\,(0.1122) & $-$0.0267\,(0.4887) & 0.0158\,(1.3546)\\
      & 5.00  && $-$0.0228\,(0.1695) & $-$0.0370\,(1.0831) & 0.0191\,(3.2483)  &&& $-$0.0426\,(0.3112) & $-$0.0428\,(1.3439) & 0.0225\,(3.7660)\\
\\
300   & 1.00  && $-$0.0006\,(0.0023) & $-$0.0048\,(0.0147) & 0.0043\,(0.0428)  &&& $-$0.0014\,(0.0040) & $-$0.0053\,(0.0187) & 0.0040\,(0.0498)\\
      & 3.00  && $-$0.0017\,(0.0209) & $-$0.0137\,(0.1302) & 0.0110\,(0.3834)  &&& $-$0.0041\,(0.0365) & $-$0.0166\,(0.1602) & 0.0123\,(0.4428)\\
      & 5.00  && $-$0.0028\,(0.0578) & $-$0.0222\,(0.3609) & 0.0171\,(1.0650)  &&& $-$0.0068\,(0.1007) & $-$0.0283\,(0.4382) & 0.0215\,(1.2277)\\
\\
500   & 1.00  && $-$0.0003\,(0.0014) & $-$0.0025\,(0.0088) & 0.0028\,(0.0258)  &&& $-$0.0007\,(0.0024) & $-$0.0016\,(0.0113) & 0.0003\,(0.0309)\\
      & 3.00  && $-$0.0006\,(0.0127) & $-$0.0064\,(0.0778) & 0.0061\,(0.2309)  &&& $-$0.0011\,(0.0221) & $-$0.0064\,(0.0981) & 0.0018\,(0.2770)\\
      & 5.00  && $-$0.0011\,(0.0354) & $-$0.0105\,(0.2159) & 0.0104\,(0.6406)  &&& $-$0.0012\,(0.0617) & $-$0.0108\,(0.2697) & 0.0023\,(0.7681)\\
\hline
\end{tabular}
\end{adjustbox}
\end{table}

\begin{table}[!ht]
\small
\centering
 \renewcommand{\arraystretch}{0.8}
 \renewcommand{\tabcolsep}{0.2cm}
\caption{
Empirical bias and MSE (in parentheses) from simulated data for the indicated ML estimators of the tobit-L$t$ 
model parameters, $n$ and $\varrho$.} \label{table:MC2}
\begin{adjustbox}{max width=\textwidth}
  \begin{tabular}{lrrrrrrrrrr}
	\hline\vspace{-0.1cm} \\
$n$  &&&  \multicolumn{3}{c}{$\varrho =0.20$} &&& \multicolumn{3}{c}{$\varrho=0.50$}  \\
   \cline{4-6} \cline{9-11} \vspace{-0.1cm} \\
& \multicolumn{1}{l}{$\phi$} && \multicolumn{1}{c}{$\widehat\phi$} & \multicolumn{1}{c}{$\widehat\beta_{0}$} & \multicolumn{1}{c}{$\widehat\beta_{1}$} &&&
\multicolumn{1}{c}{$\widehat\phi$} & \multicolumn{1}{c}{$\widehat\beta_{0}$} & \multicolumn{1}{r}{$\widehat\beta_{1}$} \\
  \hline \vspace{-0.1cm} \\
50    & 1.00  &&  $-$0.0038\,(0.0190) & $-$0.0058\,(0.1148) & 0.0085\,(0.3473) &&& 0.0056\,(0.0347) & $-$0.0217\,(0.1471) & $-$0.0181\,(0.4173)\\
      & 3.00  &&  $-$0.0118\,(0.1706) & $-$0.0113\,(1.0352) & 0.0098\,(3.1385) &&& 0.0149\,(0.3147) & $-$0.0464\,(1.2505) & $-$0.0961\,(3.6955)\\
      & 5.00  &&  $-$0.0202\,(0.4739) & $-$0.0196\,(2.8732) & 0.0287\,(8.7144) &&& 0.0252\,(0.8749) & $-$0.0715\,(3.4329) & $-$0.0308\,(9.2639)\\
\\  
100   & 1.00  &&    0.0011\,(0.0102) & $-$0.0051\,(0.0560) & 0.0050\,(0.1668)  &&& 0.0031\,(0.0185) & $-$0.0154\,(0.0719) & 0.0115\,(0.1958)\\
      & 3.00  &&    0.0023\,(0.0911) & $-$0.0111\,(0.5021) & 0.0079\,(1.4986)  &&& 0.0107\,(0.1682) & $-$0.0409\,(0.6101) & 0.0218\,(1.7335)\\
      & 5.00  &&    0.0037\,(0.2529) & $-$0.0173\,(1.3919) & 0.0197\,(4.1527)  &&& 0.0176\,(0.4673) & $-$0.0640\,(1.6827) & 0.0286\,(4.8195)\\
\\
300   & 1.00  && $-$0.0005\,(0.0033) & $-$0.0027\,(0.0181) & 0.0047\,(0.0547)  &&& 0.0003\,(0.0059) & $-$0.0069\,(0.0229) & 0.0083\,(0.0636)\\
      & 3.00  && $-$0.0018\,(0.0294) & $-$0.0073\,(0.1622) & 0.0123\,(0.4910)  &&& 0.0007\,(0.0536) & $-$0.0168\,(0.1941) & 0.0180\,(0.5621)\\
      & 5.00  && $-$0.0027\,(0.0814) & $-$0.0117\,(0.4497) & 0.0109\,(1.3623)  &&& 0.0022\,(0.1494) & $-$0.0285\,(0.5374) & 0.0292\,(1.5663)\\
\\
500   & 1.00  && $-$0.0004\,(0.0019) & $-$0.0010\,(0.0108) & 0.0011\,(0.0331)  &&& 0.0001\,(0.0035) & $-$0.0024\,(0.0134) & 0.0013\,(0.0379)\\
      & 3.00  && $-$0.0010\,(0.0173) & $-$0.0023\,(0.0972) & 0.0019\,(0.2977)  &&& 0.0005\,(0.0316) & $-$0.0037\,(0.1155) & 0.0021\,(0.3394)\\
      & 5.00  && $-$0.0018\,(0.0482) & $-$0.0038\,(0.2699) & 0.0031\,(0.8267)  &&& 0.0002\,(0.0882) & $-$0.0067\,(0.3167) & 0.0037\,(0.9394)\\
\hline
\end{tabular}
\end{adjustbox}
\end{table}

\begin{table}[!ht]
\small
\centering
 \renewcommand{\arraystretch}{0.8}
 \renewcommand{\tabcolsep}{0.2cm}
\caption{
Empirical bias and MSE (in parentheses) from simulated data for the indicated ML estimators of the tobit-LPE 
model parameters, $n$ and $\varrho$.} \label{table:MC3}
\begin{adjustbox}{max width=\textwidth}
  \begin{tabular}{lrrrrrrrrrr}
	\hline\vspace{-0.1cm} \\
$n$  &&&  \multicolumn{3}{c}{$\varrho =0.20$} &&& \multicolumn{3}{c}{$\varrho=0.50$}  \\
   \cline{4-6} \cline{9-11} \vspace{-0.1cm} \\
& \multicolumn{1}{l}{$\phi$} && \multicolumn{1}{c}{$\widehat\phi$} & \multicolumn{1}{c}{$\widehat\beta_{0}$} & \multicolumn{1}{c}{$\widehat\beta_{1}$} &&&
\multicolumn{1}{c}{$\widehat\phi$} & \multicolumn{1}{c}{$\widehat\beta_{0}$} & \multicolumn{1}{c}{$\widehat\beta_{1}$} \\
  \hline \vspace{-0.1cm} \\
50    & 1.00  && $-$0.0090\,(0.0194) & $-$0.0064\,(0.1903) & 0.0087\,(0.5807) &&& $-$0.0109\,(0.0314) & $-$0.0282\,(0.2277) & 0.0228\,(0.6642)\\
      & 3.00  && $-$0.0272\,(0.1742) & $-$0.0151\,(1.7096) & 0.0187\,(5.2207) &&& $-$0.0312\,(0.2793) & $-$0.0653\,(1.9717) & 0.0269\,(5.9557)\\
      & 5.00  && $-$0.0456\,(0.4837) & $-$0.0237\,(4.7496) & 0.0285\,(9.5076) &&& $-$0.0728\,(1.5164) & $-$0.1393\,(5.5873) & 0.0391\,(9.4605)\\
\\
100   & 1.00  && $-$0.0047\,(0.0099) & $-$0.0028\,(0.0910) & 0.0002\,(0.2815)  &&& $-$0.0037\,(0.0159) & $-$0.0149\,(0.1077) & 0.0106\,(0.3170)\\
      & 3.00  && $-$0.0061\,(0.0888) & $-$0.0081\,(0.8171) & 0.0142\,(2.5290)  &&& $-$0.0122\,(0.1427) & $-$0.0339\,(0.9188) & 0.0174\,(2.7931)\\
      & 5.00  && $-$0.0102\,(0.2469) & $-$0.0135\,(2.2718) & 0.0048\,(7.0339)  &&& $-$0.0205\,(0.3969) & $-$0.0537\,(2.5311) & 0.0170\,(7.7414)\\
\\
300   & 1.00  && $-$0.0021\,(0.0032) & $-$0.0019\,(0.0295) & 0.0002\,(0.0865)  &&& $-$0.0010\,(0.0052) & $-$0.0063\,(0.0341) & 0.0043\,(0.0956)\\
      & 3.00  && $-$0.0026\,(0.0287) & $-$0.0051\,(0.2653) & 0.0030\,(0.7783)  &&& $-$0.0069\,(0.0468) & $-$0.0172\,(0.2901) & 0.0130\,(0.8346)\\
      & 5.00  && $-$0.0047\,(0.0798) & $-$0.0078\,(0.7366) & 0.0021\,(2.1619)  &&& $-$0.0097\,(0.1299) & $-$0.0253\,(0.7998) & 0.0047\,(2.3190)\\
\\
500   & 1.00  && $-$0.0008\,(0.0019) & $-$0.0017\,(0.0177) & 0.0001\,(0.0525)  &&& $-$0.0003\,(0.0031) & $-$0.0005\,(0.0204) & 0.0033\,(0.0574)\\
      & 3.00  && $-$0.0021\,(0.0176) & $-$0.0047\,(0.1593) & 0.0018\,(0.4733)  &&& $-$0.0017\,(0.0280) & $-$0.0033\,(0.1730) & 0.0030\,(0.5050)\\
      & 5.00  && $-$0.0011\,(0.0176) & $-$0.0051\,(0.1593) & 0.0014\,(0.4733)  &&& $-$0.0079\,(0.0775) & $-$0.0162\,(0.4754) & 0.0031\,(1.3952)\\
\hline
\end{tabular}
\end{adjustbox}
\end{table}

\newpage

\subsection{Statistical tests}\label{test:mc}
We now present a MC simulation study to evaluate and compare the performance of the LR and {GR} tests. We consider again the following models: tobit-LN, tobit-L$t$ and tobit-LPE. 
The simulation scenario considers:  sample size $n \in \{50, 100, 300, 500\}$, scale parameter $\phi=3.00$, 
extra parameter $\xi_{1}=0.5$ (tobit-LPE), $\xi_{1}=5$ (tobit-L$t$), censoring proportion $\varrho=m/n \in \{0.3,0.5\}$, 
with 5,000 {MC} replications for each sample size. We consider as data generating 
process the model
\begin{equation*}
Y_{i} =
\begin{cases}
   \gamma, &  Y_{i}^{\ast} \leq  \gamma,\,\, i=1,\ldots, m,\\
  Y_{i}^{\ast}= \beta_{0} + \beta_{1}x_{1i}+\beta_{2}x_{2i}+\beta_{3}x_{3i}+\beta_{4}x_{4i}+\varepsilon_{i}, & Y_{i}^{\ast}> \gamma,\,\, i=m+1,\ldots,n,\\
\end{cases}
\end{equation*}
where $\varepsilon_{i}$ is as in \eqref{eq:tobit-logsym}, with $\beta_{0}=1.0$, $\beta_{1}=1.5$, $\beta_{2}=0.5$, $\beta_{3}=0.8$ and
$\beta_{4}\in \{-1.00, -0.75, -0.25$ $0.00, 0.25,0.75,1.00\}$. The covariate values were taken as random draws from the 
U(0,1) distribution. The interest lies in testing ${\cal{H}}_{0}:{\beta}_{4}=0$ against ${\cal{H}}_{1}:{\beta}_{4}\neq0$.

Tables \ref{table:PWTEST1}-\ref{table:PWTEST3} present the simulation results regarding the powers of the tests, namely, their capacity to identify a false null hypothesis. 
Note, however, that we also consider the case where the null hypothesis is true ($\beta_{4}=0.00$ in the data generation). From Tables \ref{table:PWTEST1}-\ref{table:PWTEST3}, we observe that the power associated with the LR and GR tests increases as a function of the sample size, as expected. We also observe that the power of the tests decreases when the censoring proportion increases. In general, the results show that both tests have similar power.

\begin{table}[!ht]
\small
\centering
 \renewcommand{\arraystretch}{0.7}
 \renewcommand{\tabcolsep}{0.1cm}
\caption{{Power study ($\%$) for different values of $\beta_{4}$ and models (nominal level = 1\%).}} \label{table:PWTEST1}
\begin{adjustbox}{max width=\textwidth}
  \begin{tabular}{llrcccccccrrrrrrrrrr}
	\hline\vspace{-0.1cm} \\
    &    &             &  \multicolumn{5}{c}{tobit-LN}                                               && \multicolumn{5}{c}{tobit-L$t$}                                            && \multicolumn{5}{c}{tobit-LPE}      \\ \cline{4-8} \cline{10-14} \cline{16-20} \vspace{-0.15cm}\\ 
    &    &             &  \multicolumn{2}{c}{$\varrho =0.20$} && \multicolumn{2}{c}{$\varrho=0.50$}  && \multicolumn{2}{c}{$\varrho =0.20$} && \multicolumn{2}{c}{$\varrho=0.50$} && \multicolumn{2}{c}{$\varrho =0.20$} && \multicolumn{2}{c}{$\varrho=0.50$}  \\ \cline{4-5} \cline{7-8} \cline{10-11} \cline{13-14}  \cline{16-17}  \cline{19-20}\\  
    &    & \multicolumn{1}{c}{$\beta_{4}$} & \multicolumn{1}{c}{LR} & \multicolumn{1}{c}{{GR}}  && \multicolumn{1}{c}{LR} & \multicolumn{1}{c}{{GR}}  && \multicolumn{1}{c}{LR} & \multicolumn{1}{c}{{GR}} && \multicolumn{1}{c}{LR} & \multicolumn{1}{c}{{GR}} && \multicolumn{1}{c}{LR} & \multicolumn{1}{c}{{GR}} && \multicolumn{1}{c}{LR} & \multicolumn{1}{c}{{GR}}\\
  \hline \vspace{-0.1cm} \\
$n$ &50  & $-$1.00     &  4.04 & 3.18  && 3.38  &  2.62 &&  4.00 &  3.16 &&  3.54 &  2.92 &&  4.26 &  3.38 &&  3.66 & 3.18\\
    &    & $-$0.75     &  2.90 & 2.38  && 2.56  &  1.94 &&  2.88 &  2.48 &&  2.82 &  2.28 &&  3.24 &  2.48 &&  2.82 & 2.52\\
    &    & $-$0.25     &  1.86 & 1.34  && 1.58  &  1.20 &&  2.28 &  1.60 &&  2.12 &  1.62 &&  2.44 &  1.72 &&  2.28 & 1.92\\
    &    &    0.00     &  1.60 & 1.24  && 1.46  &  1.16 &&  1.88 &  1.40 &&  1.86 &  1.48 &&  2.04 &  1.62 &&  2.08 & 1.60\\
    &    &    0.25     &  1.76 & 1.42  && 1.74  &  1.24 &&  2.02 &  1.62 &&  1.92 &  1.50 &&  2.36 &  1.82 &&  2.20 & 1.78\\
    &    &    0.75     &  2.74 & 2.06  && 2.56  &  2.06 &&  3.00 &  2.14 &&  2.58 &  2.28 &&  3.16 &  2.44 &&  2.84 & 2.38\\
    &    &    1.00     &  3.48 & 2.80  && 3.36  &  2.78 &&  3.62 &  3.06 &&  3.40 &  2.96 &&  3.98 &  3.10 &&  4.06 & 3.20\\[-1.3ex]
\\
    &100  & $-$1.00    &  5.60 &  5.08 && 4.84  &  4.34 &&  5.82 &  5.36 &&  5.28 &  4.80 &&  6.50 &  5.60 &&  5.60 & 5.08\\
    &    & $-$0.75     &  3.52 &  3.30 && 3.28  &  2.96 &&  3.62 &  3.30 &&  3.58 &  3.28 &&  4.58 &  3.84 &&  3.84 & 3.50\\
    &    & $-$0.25     &  1.42 &  1.24 && 1.44  &  1.26 &&  1.76 &  1.54 &&  1.72 &  1.62 &&  1.96 &  1.70 &&  1.98 & 1.84\\
    &    &    0.00     &  1.12 &  1.02 && 1.34  &  1.14 &&  1.38 &  1.26 &&  1.54 &  1.44 &&  1.68 &  1.42 &&  1.78 & 1.64\\
    &    &    0.25     &  1.40 &  1.24 && 1.46  &  1.36 &&  1.82 &  1.50 &&  1.72 &  1.54 &&  1.98 &  1.56 &&  1.84 & 1.64\\
    &    &    0.75     &  2.90 &  2.60 && 2.84  &  2.50 &&  3.58 &  3.16 &&  3.14 &  2.84 &&  3.82 &  3.40 &&  3.40 & 3.30\\
    &    &    1.00     &  4.72 &  4.18 && 4.20  &  3.82 &&  4.98 &  4.62 &&  4.62 &  4.36 &&  5.80 &  5.26 &&  4.98 & 4.62\\[-1.3ex]   
\\
    &300  & $-$1.00    & 15.52 & 15.20 && 12.90 & 12.52 && 16.54 & 16.20 && 13.78 & 13.66 && 17.34 & 16.90 && 14.62 & 14.50\\
    &    & $-$0.75     &  7.90 &  7.60 &&  6.56 &  6.44 &&  8.48 &  8.22 &&  7.24 &  7.00 &&  9.62 &  9.18 &&  7.84 & 7.62\\
    &    & $-$0.25     &  1.64 &  1.60 &&  1.52 &  1.42 &&  1.84 &  1.76 &&  1.82 &  1.72 &&  2.16 &  2.04 &&  1.84 & 1.84\\
    &    &    0.00     &  1.10 &  0.98 &&  1.18 &  1.12 &&  1.44 &  1.38 &&  1.41 &  1.40 &&  1.70 &  1.62 &&  1.80 & 1.64\\
    &    &    0.25     &  1.78 &  1.74 &&  1.80 &  1.78 &&  1.88 &  1.74 &&  2.08 &  1.98 &&  2.14 &  2.02 &&  2.20 & 2.18\\
    &    &    0.75     &  8.54 &  8.14 &&  7.42 &  7.24 &&  9.00 &  8.70 &&  8.12 &  7.84 &&  9.84 &  9.32 &&  8.32 & 7.92\\
    &    &    1.00     & 16.34 & 16.06 && 13.68 & 13.14 && 16.88 & 16.46 && 14.08 & 13.78 && 17.54 & 16.96 && 14.74 & 14.30\\[-1.3ex]    
\\
    &500  & $-$1.00    & 29.58 & 29.12 && 23.40 & 23.04 && 29.34 & 29.16 && 23.74 & 23.46 && 29.90 & 29.50 && 24.28 & 24.04\\
    &    & $-$0.75     & 13.72 & 13.56 && 11.30 & 11.12 && 14.08 & 13.74 && 11.90 & 11.72 && 15.20 & 14.94 && 12.74 & 12.42\\
    &    & $-$0.25     &  2.00 &  1.94 && 1.94  &  1.86 &&  2.44 &  2.40 &&  2.34 &  2.32 &&  3.00 &  3.02 &&  2.62 & 2.60\\
    &    &    0.00     &  1.08 &  1.08 && 1.04  &  1.02 &&  1.30 &  1.22 &&  1.16 &  1.10 &&  1.86 &  1.80 &&  1.64 & 1.62\\
    &    &    0.25     &  2.02 &  1.98 && 1.90  &  1.82 &&  2.46 &  2.36 &&  2.24 &  2.22 &&  2.80 &  2.60 &&  2.38 & 2.40\\
    &    &    0.75     & 15.24 & 15.16 && 12.82 & 12.52 && 15.44 & 15.32 && 12.84 & 12.80 && 16.38 & 16.04 && 13.46 & 13.20\\
    &    &    1.00     & 29.84 & 29.52 && 24.56 & 24.20 && 30.26 & 29.98 && 25.12 & 24.88 && 30.90 & 30.16 && 25.12 & 25.00\\
    
\hline
\end{tabular}
\end{adjustbox}
\end{table}

\begin{table}[!ht]
\small
\centering
 \renewcommand{\arraystretch}{0.7}
 \renewcommand{\tabcolsep}{0.1cm}
\caption{{Power study ($\%$) for different values of $\beta_{4}$ and models (nominal level = 5\%).}} \label{table:PWTEST2}
\begin{adjustbox}{max width=\textwidth}
  \begin{tabular}{llrcccccccrrrrrrrrrr}
	\hline\vspace{-0.1cm} \\
    &    &             &  \multicolumn{5}{c}{tobit-LN}                                               && \multicolumn{5}{c}{tobit-L$t$}                                            && \multicolumn{5}{c}{tobit-LPE}      \\ \cline{4-8} \cline{10-14} \cline{16-20} \vspace{-0.15cm}\\ 
    &    &             &  \multicolumn{2}{c}{$\varrho =0.20$} && \multicolumn{2}{c}{$\varrho=0.50$}  && \multicolumn{2}{c}{$\varrho =0.20$} && \multicolumn{2}{c}{$\varrho=0.50$} && \multicolumn{2}{c}{$\varrho =0.20$} && \multicolumn{2}{c}{$\varrho=0.50$}  \\ \cline{4-5} \cline{7-8} \cline{10-11} \cline{13-14}  \cline{16-17}  \cline{19-20}\\  
    &    & \multicolumn{1}{c}{$\beta_{4}$} & \multicolumn{1}{c}{LR} & \multicolumn{1}{c}{{GR}}  && \multicolumn{1}{c}{LR} & \multicolumn{1}{c}{{GR}}  && \multicolumn{1}{c}{LR} & \multicolumn{1}{c}{{GR}} && \multicolumn{1}{c}{LR} & \multicolumn{1}{c}{{GR}} && \multicolumn{1}{c}{LR} & \multicolumn{1}{c}{{GR}} && \multicolumn{1}{c}{LR} & \multicolumn{1}{c}{{GR}}\\
  \hline \vspace{-0.1cm} \\
$n$ &50  & $-$1.00     & 11.58 & 10.76 && 10.98 & 10.06 && 12.82 & 11.96 && 11.76 & 10.92 && 13.66 & 12.46 && 12.40 & 11.52\\
    &    & $-$0.75     & 9.84  &  9.14 && 9.22  & 8.54  && 10.52 &  9.94 &&  9.92 &  9.30 && 11.50 & 10.24 && 10.58 & 9.82\\
    &    & $-$0.25     & 7.58  &  6.84 && 7.06  & 6.48  &&  7.74 &  7.12 &&  7.90 &  6.92 &&  8.56 &  7.60 &&  8.38 & 7.86\\
    &    &    0.00     & 7.08  & 6.66  && 6.78  & 6.00  &&  7.60 &  6.74 &&  7.12 &  6.44 &&  8.04 &  7.16 &&  7.82 & 7.14 \\
    &    &    0.25     & 7.42  & 6.76  && 6.74  & 6.16  &&  7.72 &  7.22 &&  7.22 &  6.80 &&  8.48 &  7.42 &&  8.16 & 7.30\\
    &    &    0.75     & 8.90  & 8.26  && 8.60  & 7.82  &&  9.60 &  8.84 &&  9.04 &  8.48 && 10.94 &  9.66 && 10.08 & 9.30\\
    &    &    1.00     & 10.66 & 9.92  && 9.90  & 9.30  && 11.50 & 10.42 && 10.76 & 10.04 && 12.28 & 11.44 && 11.72 & 10.88\\[-1.3ex]
\\
    &100  & $-$1.00    & 16.30 & 15.76 && 14.32 & 14.02 && 16.88 & 16.30 && 15.06 & 14.68 && 17.78 & 17.28 && 15.44 & 15.18\\
    &    & $-$0.75     & 11.36 & 11.04 && 10.94 & 10.64 && 12.44 & 11.82 && 11.52 & 11.12 && 13.34 & 12.54 && 12.36 & 11.86\\
    &    & $-$0.25     & 6.72  & 6.66  && 6.64  & 6.28  && 7.22	 & 6.74  &&  6.92 &  6.58 &&  8.02 &  7.56 &&  7.44 & 7.20\\
    &    &    0.00     & 5.36  & 5.14  && 5.56  & 5.30  &&  6.30 &  5.90 &&  6.30 &  5.98 &&  7.36 &  6.68 &&  6.26 & 6.18\\
    &    &    0.25     & 5.70  & 5.42  && 5.78  & 5.54  &&  6.50 &  6.26 &&  6.72 &  6.32 &&  7.44 &  7.00 &&  7.10 & 6.70\\
    &    &    0.75     & 10.94 & 10.54 && 10.06 & 9.60  && 11.84 & 11.28 && 10.72 & 10.48 && 12.64 & 11.82 && 11.70 & 11.30\\
    &    &    1.00     & 15.42 & 14.86 && 13.46 & 13.02 && 16.32 & 15.64 && 14.14 & 13.78 && 16.82 & 15.92 && 14.98 & 14.54\\[-1.3ex]   
\\
    &300  & $-$1.00    & 34.22 & 33.94 && 30.08 & 29.80 && 35.06 & 34.84 && 30.84 & 30.64 && 34.98 & 34.70 && 31.30 & 31.30\\
    &    & $-$0.75     & 21.94 & 21.78 && 19.16 & 18.90 && 22.56 & 22.40 && 20.20 & 20.12 && 23.38 & 23.14 && 20.68 & 20.72\\
    &    & $-$0.25     & 6.46  & 6.22  && 6.74  & 6.66  &&  7.40 &  7.28 &&  7.18 &  7.12 &&  8.62 &  8.50 &&  7.80 & 7.70\\
    &    &    0.00     & 5.30  & 5.28  && 5.28  & 5.22  &&  5.62 &  5.54 &&  5.68 &  5.58 &&  6.18 &  6.10 &&  6.18 & 6.12\\
    &    &    0.25     & 6.78  & 6.70  && 7.06  & 6.92  &&  7.60 &  7.40 &&  7.68 &  7.54 &&  8.60 &  8.50 &&  8.08 & 8.00\\
    &    &    0.75     & 22.56 & 22.46 && 19.96 & 19.62 && 22.80 & 22.58 && 20.12 & 19.90 && 24.04 & 23.70 && 20.80 & 20.42\\
    &    &    1.00     & 35.04 & 34.60 && 31.22 & 30.92 && 35.38 & 35.22 && 31.24 & 31.02 && 36.12 & 36.08 && 32.18 & 32.08\\[-1.3ex]    
\\
    &500  & $-$1.00    & 53.32 & 53.18 && 47.88 & 47.68 && 52.94 & 52.74 && 47.68 & 47.60 && 52.50 & 52.36 && 47.38 & 47.38\\
    &    & $-$0.75     & 33.12 & 32.96 && 28.62 & 28.50 && 33.08 & 32.96 && 28.70 & 28.46 && 33.64 & 33.38 && 29.34 & 29.14\\
    &    & $-$0.25     & 7.82  & 7.78  && 7.36  & 7.30  &&  8.64 &  8.54 &&  8.24 &  8.14 &&  9.46 &  9.42 &&  8.58 & 8.60\\
    &    &    0.00     & 5.40  & 5.40  && 5.22  & 5.10  &&  5.72 &  5.64 &&  5.72 &  5.62 &&  6.68 &  6.50 &&  6.26 & 6.22\\
    &    &    0.25     & 8.16  & 8.12  && 7.98  & 7.78  &&  8.90 &  8.86 &&  8.52 &  8.42 &&  9.54 &  9.36 &&  9.16 & 9.04\\
    &    &    0.75     & 33.86 & 33.78 && 29.16 & 29.02 && 34.42 & 34.26 && 30.10 & 29.94 && 34.48 & 34.54 && 29.90 & 29.88\\
    &    &    1.00     & 53.10 & 53.06 && 46.98 & 46.78 && 52.96 & 52.94 && 47.24 & 47.08 && 52.70 & 52.58 && 47.44 & 47.26\\
    
\hline
\end{tabular}
\end{adjustbox}
\end{table}

\begin{table}[!ht]
\small
\centering
 \renewcommand{\arraystretch}{0.7}
 \renewcommand{\tabcolsep}{0.1cm}
\caption{{Power study ($\%$) for different values of $\beta_{4}$ and models (nominal level = 10\%).}} \label{table:PWTEST3}
\begin{adjustbox}{max width=\textwidth}
  \begin{tabular}{llrcccccccrrrrrrrrrr}
	\hline\vspace{-0.1cm} \\
    &    &             &  \multicolumn{5}{c}{tobit-LN}                                               && \multicolumn{5}{c}{tobit-L$t$}                                            && \multicolumn{5}{c}{tobit-LPE}      \\ \cline{4-8} \cline{10-14} \cline{16-20} \vspace{-0.15cm}\\ 
    &    &             &  \multicolumn{2}{c}{$\varrho =0.20$} && \multicolumn{2}{c}{$\varrho=0.50$}  && \multicolumn{2}{c}{$\varrho =0.20$} && \multicolumn{2}{c}{$\varrho=0.50$} && \multicolumn{2}{c}{$\varrho =0.20$} && \multicolumn{2}{c}{$\varrho=0.50$}  \\ \cline{4-5} \cline{7-8} \cline{10-11} \cline{13-14}  \cline{16-17}  \cline{19-20}\\  
    &    & \multicolumn{1}{c}{$\beta_{4}$} & \multicolumn{1}{c}{LR} & \multicolumn{1}{c}{{GR}}  && \multicolumn{1}{c}{LR} & \multicolumn{1}{c}{{GR}}  && \multicolumn{1}{c}{LR} & \multicolumn{1}{c}{{GR}} && \multicolumn{1}{c}{LR} & \multicolumn{1}{c}{{GR}} && \multicolumn{1}{c}{LR} & \multicolumn{1}{c}{{GR}} && \multicolumn{1}{c}{LR} & \multicolumn{1}{c}{{GR}}\\
  \hline \vspace{-0.1cm} \\
$n$ &50  & $-$1.00     & 18.92 & 18.28 && 18.06 & 17.54 && 20.06 & 19.46 && 19.52 & 18.78 && 21.02 & 19.86 && 20.00 & 19.64\\
    &    & $-$0.75     & 16.50 & 16.00 && 16.24 & 15.52 && 17.46 & 16.76 && 17.40 & 16.68 && 18.40 & 17.44 && 17.98 & 17.30\\
    &    & $-$0.25     & 13.42 & 12.94 && 12.98 & 12.36 && 14.54 & 13.94 && 13.54 & 12.96 && 15.76 & 14.40 && 14.68 & 13.92\\
    &    &    0.00     & 12.68 & 12.20 && 12.58 & 11.92 && 13.66 & 13.20 && 13.22 & 12.36 && 15.20 & 14.14 && 14.26 & 13.46\\
    &    &    0.25     & 12.88 & 12.42 && 12.68 & 12.02 && 14.00 & 13.22 && 13.38 & 12.84 && 14.88 & 14.28 && 14.42 & 13.54\\
    &    &    0.75     & 15.54 & 14.88 && 14.48 & 13.76 && 16.50 & 15.68 && 15.72 & 15.18 && 17.40 & 16.18 && 16.76 & 16.20\\
    &    &    1.00     & 18.06 & 17.44 && 16.66 & 16.02 && 19.18 & 18.56 && 17.92 & 17.24 && 20.20 & 18.90 && 18.60 & 17.88\\[-1.3ex]
\\
    &100  & $-$1.00    & 24.26 & 23.70 && 22.84 & 22.54 && 25.08 & 24.90 && 23.10 & 22.86 && 25.98 & 25.12 && 23.62 & 23.40\\
    &    & $-$0.75     & 19.06 & 18.82 && 17.86 & 17.28 && 20.10 & 19.80 && 18.50 & 18.32 && 20.80 & 20.24 && 18.74 & 18.76\\
    &    & $-$0.25     & 11.98 & 11.80 && 11.90 & 11.62 && 13.00 & 12.74 && 12.68 & 12.48 && 14.46 & 13.76 && 13.84 & 13.32\\
    &    &    0.00     & 11.22 & 10.90 && 11.50 & 11.20 && 11.70 & 11.52 && 12.14 & 12.00 && 12.88 & 12.62 && 12.82 & 12.58\\
    &    &    0.25     & 11.80 & 11.50 && 11.88 & 11.52 && 12.40 & 12.16 && 12.66 & 12.36 && 14.10 & 13.42 && 13.78 & 13.58\\
    &    &    0.75     & 18.36 & 17.94 && 17.56 & 17.04 && 19.94 & 19.48 && 18.06 & 17.90 && 20.94 & 20.18 && 18.82 & 18.52\\
    &    &    1.00     & 24.56 & 24.02 && 22.48 & 21.92 && 25.38 & 25.00 && 23.06 & 22.64 && 26.48 & 26.00 && 23.70 & 23.20\\[-1.3ex]   
\\
    &300  & $-$1.00    & 46.70 & 46.66 && 42.04 & 41.90 && 46.56 & 46.48 && 42.02 & 41.98 && 46.52 & 46.50 && 42.78 & 42.64\\
    &    & $-$0.75     & 32.12 & 31.98 && 29.26 & 29.06 && 32.90 & 32.74 && 30.24 & 30.16 && 33.34 & 32.94 && 30.60 & 30.52\\
    &    & $-$0.25     & 12.76 & 12.68 && 11.78 & 11.76 && 13.80 & 13.66 && 13.68 & 13.62 && 15.06 & 14.66 && 14.62 & 14.54\\
    &    &    0.00     & 9.58  & 9.54  && 10.34 & 10.26 && 10.74 & 10.68 && 11.28 & 11.20 && 12.62 & 12.62 && 12.82 & 12.58\\
    &    &    0.25     & 13.26 & 13.22 && 12.68 & 12.58 && 13.76 & 13.62 && 13.28 & 13.26 && 14.60 & 14.48 && 14.48 & 14.22\\
    &    &    0.75     & 32.60 & 32.46 && 29.94 & 29.74 && 33.40 & 33.30 && 30.38 & 30.18 && 34.14 & 33.68 && 31.04 & 30.70\\
    &    &    1.00     & 47.20 & 47.08 && 43.34 & 43.04 && 47.40 & 47.28 && 43.36 & 43.24 && 47.16 & 47.04 && 43.86 & 43.54\\[-1.3ex]    
\\
    &500  & $-$1.00    & 64.84 & 64.78 && 59.90 & 59.82 && 65.06 & 65.02 && 59.90 & 59.78 && 64.18 & 63.96 && 59.92 & 59.88\\
    &    & $-$0.75     & 45.82 & 45.74 && 41.26 & 41.12 && 45.40 & 45.30 && 41.22 & 41.14 && 46.02 & 45.92 && 41.28 & 41.20\\
    &    & $-$0.25     & 13.62 & 13.56 && 13.04 & 12.92 && 14.26 & 14.24 && 14.14 & 13.98 && 15.78 & 15.58 && 14.92 & 14.76\\
    &    &    0.00     & 10.66 & 10.04 && 10.76 & 10.68 && 10.84 & 10.84 && 10.97 & 10.90 && 12.32 & 12.22 && 12.04 & 11.90\\
    &    &    0.25     & 14.80 & 14.68 && 14.12 & 14.04 && 15.34 & 15.30 && 15.12 & 15.08 && 16.66 & 16.46 && 15.70 & 15.70\\
    &    &    0.75     & 45.98 & 45.90 && 41.46 & 41.40 && 45.46 & 45.36 && 41.16 & 41.08 && 45.82 & 45.86 && 41.84 & 41.84\\
    &    &    1.00     & 65.68 & 65.66 && 60.16 & 60.12 && 65.58 & 65.54 && 60.14 & 60.00 && 65.64 & 65.24 && 59.74 & 59.86\\
    
\hline
\end{tabular}
\end{adjustbox}
\end{table}

\section{Application}\label{sec:5}

Tobit-log-symmetric models are now used to analyse a data set from a case-study of measles vaccines, corresponding to antibody concentration levels (response variable, $Y_{i}$)
collected from 330 children at 12 months of age; see \cite{mh:95}. In the measurement of antibody concentration by 
quantitative assays, there is always a concentration value, $\gamma$ say, below which an exact measurement 
cannot be computed, independently of the employed technique. Then, this value $\gamma$ can be used to substitute a value for the 
censored observation. In the measles vaccine data, the value of $\gamma$ was 0.1 international units (IU) or $-$2.306 in logarithm scale. 
It was verified that 86 (26.1\%) of the observations fell below $\gamma$ and then were recorded as 0.1. The covariates considered in the study were: 
$x_{i1}$ is the type of vaccine used (0 if Schwartz and 1 if Edmonston-Zagreb); $x_{i2}$ is the level of the dosage 
(0 if medium and 1 if high); and $x_{i3}$ is the gender where 0 is male and 1 is female.

Table~\ref{tab:estdesc} reports descriptive statistics of the observed antibody concentration levels, including the 
median (MD), mean ($\overline{y}$), standard deviation (SD), coefficient of variation (CV), skewness (CS) and kurtosis 
(CK), and minimum (${y}_{(1)}$) and maximum (${y}_{(n)}$) values. From this table, note the right skewed nature and high 
kurtosis level of the data distribution.

\begin{table}[!ht]
\centering
\caption{Summary statistics for the measles vaccine data.}
\label{tab:estdesc}
\begin{tabular}{ccccccccccccccccc}
\hline
MD    &   $\overline{y}$   & MD       &  SD     &  CV       &  CS   &  CK    & ${y}_{(1)}$ & ${y}_{(n)}$ & $n$  \\
\hline
0.4   &  1.20              & 0.40     & 2.10   & 174.74\%  &  3.46  & 14.37 &     0.10     &    15.47   & 330 \\
\hline
\end{tabular}
\end{table}

\begin{figure}[!ht]
\centering
\psfrag{0.000}[c][c]{\scriptsize{.0}}
\psfrag{0.001}[c][c]{\scriptsize{.001}}
\psfrag{0.002}[c][c]{\scriptsize{.002}}
\psfrag{0.003}[c][c]{\scriptsize{.003}}
\psfrag{0.004}[c][c]{\scriptsize{.004}}
\psfrag{0.005}[c][c]{\scriptsize{.005}}
\psfrag{0.006}[c][c]{\scriptsize{.006}}
\psfrag{0.0}[c][c]{\scriptsize{0.0}}
\psfrag{0.1}[c][c]{\scriptsize{0.1}}
\psfrag{0.2}[c][c]{\scriptsize{0.2}}
\psfrag{0.3}[c][c]{\scriptsize{0.3}}
\psfrag{0.4}[c][c]{\scriptsize{0.4}}

\psfrag{0.5}[c][c]{\scriptsize{0.5}}
\psfrag{0.6}[c][c]{\scriptsize{0.6}}
\psfrag{0.7}[c][c]{\scriptsize{0.7}}
\psfrag{0.8}[c][c]{\scriptsize{0.8}}
\psfrag{1.0}[c][c]{\scriptsize{1.0}}
\psfrag{5}[c][c]{\scriptsize{5}}
\psfrag{10}[c][c]{\scriptsize{10}}
\psfrag{15}[c][c]{\scriptsize{15}}
\psfrag{20}[c][c]{\scriptsize{20}}
\psfrag{0}[c][c]{\scriptsize{0}}
\psfrag{5}[c][c]{\scriptsize{5}}
\psfrag{10}[c][c]{\scriptsize{10}}
\psfrag{15}[c][c]{\scriptsize{15}}
\psfrag{200}[c][c]{\scriptsize{200}}
\psfrag{400}[c][c]{\scriptsize{400}}
\psfrag{600}[c][c]{\scriptsize{600}}
\psfrag{800}[c][c]{\scriptsize{800}}
\psfrag{1000}[c][c]{\scriptsize{1000}}
\psfrag{dp}[c][c]{\scriptsize{antibody concentration}}
\psfrag{de}[c][c]{\scriptsize{PDF}}
\psfrag{aa}[c][c]{\scriptsize{usual boxplot}}
\psfrag{ad}[c][c]{\scriptsize{adjusted boxplot}}
\subfigure[Histogram]{\includegraphics[height=5cm,width=5cm]{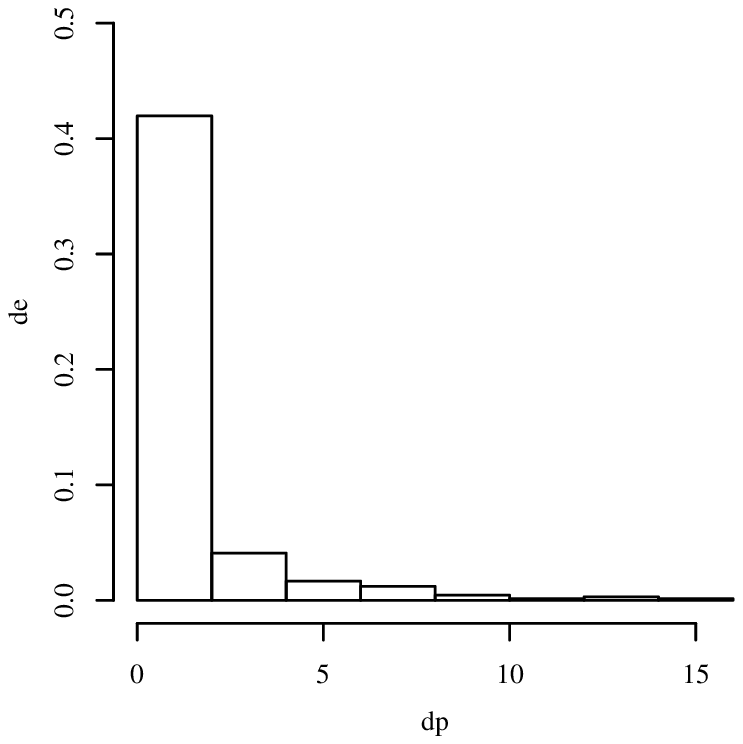}}
\subfigure[Boxplots]{\includegraphics[height=5cm,width=5cm]{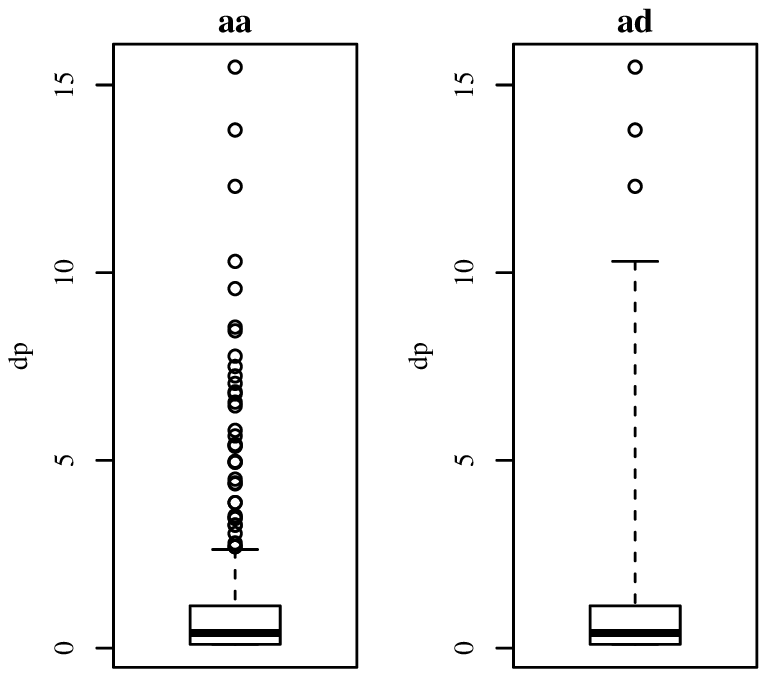}}
\caption{Histogram and boxplots for the measles vaccine data.}
  \label{fig:exploratory}
\end{figure}

Figure~\ref{fig:exploratory} presents the histogram and boxplots for the measles vaccine data. Note that 
the skewness observed in Table~\ref{tab:estdesc} is confirmed by the histogram shown in Figure~\ref{fig:exploratory}(a). 
The adjusted boxplot for the measles vaccine data indicates that some potential outliers identified by the usual boxplot 
are not outliers; see Figure~\ref{fig:exploratory}(b). The adjusted boxplot is used when the data is skew distributed; see \cite{hvv:08}.

We now analyse the measles vaccine data using the tobit-log-symmetric model, which can be 
written as
\begin{equation*}
Y_{i} =
\begin{cases}
   0.1, &  Y_{i}^{\ast} \leq  0.1,\,\, i=1,\ldots, 85,\\
  Y_{i}^{\ast}= \beta_{0} + \beta_{1}x_{i1}+\beta_{2}x_{2i}+\beta_{3}x_{3i}+\varepsilon_{i}, & Y_{i}^{\ast}>  0.1,\,\, i=86,\ldots,330,\\
\end{cases}
\end{equation*}
where $\varepsilon_{i} \stackrel{\textrm{IID}}\sim\textrm{S}(0,1,g)$. In addition to the tobit-log-symmetric models studied in 
the simulation study, we also consider the tobit-Birnbaum-Saunders (tobit-BS) and tobit-Birnbaum-Saunders-$t$ (tobit-BS-$t$) models. 

Table \ref{tab:mleresults} reports the ML estimates, computed by the BFGS quasi-Newton method, 
SEs and Akaike (AIC) and Bayesian information (BIC) criteria. For comparison, the results of the classical tobit-NO model \citep{t:58} showed 
in Equation \eqref{eq:tobitnormal}, are given as well. From Table \ref{tab:mleresults}, note that, all the tobit-log-symmetric 
models provide better adjustments compared to the tobit-NO model based on the values of AIC and BIC. Particularly, the tobit-LN 
has the lowest AIC and BIC values. 

\begin{figure}[!ht]
\centering
\psfrag{R}[c]{\scriptsize{empirical quantile}}
\psfrag{Q}[c]{\scriptsize{theoretical quantile}}
\psfrag{0}[c][c]{\scriptsize{0}}
\psfrag{1}[c][c]{\scriptsize{1}}
\psfrag{2}[c][c]{\scriptsize{2}}
\psfrag{3}[c][c]{\scriptsize{3}}
\psfrag{4}[c][c]{\scriptsize{4}}
\psfrag{5}[c][c]{\scriptsize{5}}
\psfrag{6}[c][c]{\scriptsize{6}}
\psfrag{7}[c][c]{\scriptsize{7}}
\psfrag{8}[c][c]{\scriptsize{8}}
\psfrag{9}[c][c]{\scriptsize{9}}
\psfrag{10}[c][c]{\scriptsize{10}}
\psfrag{15}[c][c]{\scriptsize{15}}
\psfrag{-0.002}[c][c]{\scriptsize{$-$.002}}
\psfrag{0.000}[c][c]{\scriptsize{.0}}
\psfrag{0.002}[c][c]{\scriptsize{.002}}
\psfrag{0.004}[c][c]{\scriptsize{.004}}
\psfrag{0.006}[c][c]{\scriptsize{.006}}
\psfrag{0.008}[c][c]{\scriptsize{.008}}
\psfrag{0.010}[c][c]{\scriptsize{.010}}
\psfrag{0.0}[c][c]{\scriptsize{0.0}}
\psfrag{0.1}[c][c]{\scriptsize{0.1}}
\psfrag{0.2}[c][c]{\scriptsize{0.2}}
\psfrag{0.3}[c][c]{\scriptsize{0.3}}
\psfrag{0.4}[c][c]{\scriptsize{0.4}}
\psfrag{0.5}[c][c]{\scriptsize{0.5}}
\psfrag{0.6}[c][c]{\scriptsize{0.6}}
\psfrag{0.7}[c][c]{\scriptsize{0.7}}
\psfrag{0.8}[c][c]{\scriptsize{0.8}}
\psfrag{1.0}[c][c]{\scriptsize{1.0}}
\psfrag{0}[c][c]{\scriptsize{0}}
\psfrag{50}[c][c]{\scriptsize{50}}
\psfrag{100}[c][c]{\scriptsize{100}}
\psfrag{150}[c][c]{\scriptsize{150}}
\psfrag{200}[c][c]{\scriptsize{200}}
\psfrag{250}[c][c]{\scriptsize{250}}
\psfrag{300}[c][c]{\scriptsize{300}}
\psfrag{328}[c]{\scriptsize{$328$}}
\psfrag{330}[l]{\scriptsize{$330$}}
\psfrag{in}[c]{\scriptsize{index}}
\psfrag{GCDn}{\scriptsize{GCD($\bm\theta$)}}
\subfigure[tobit-NO]{\includegraphics[height=4cm,width=4cm]{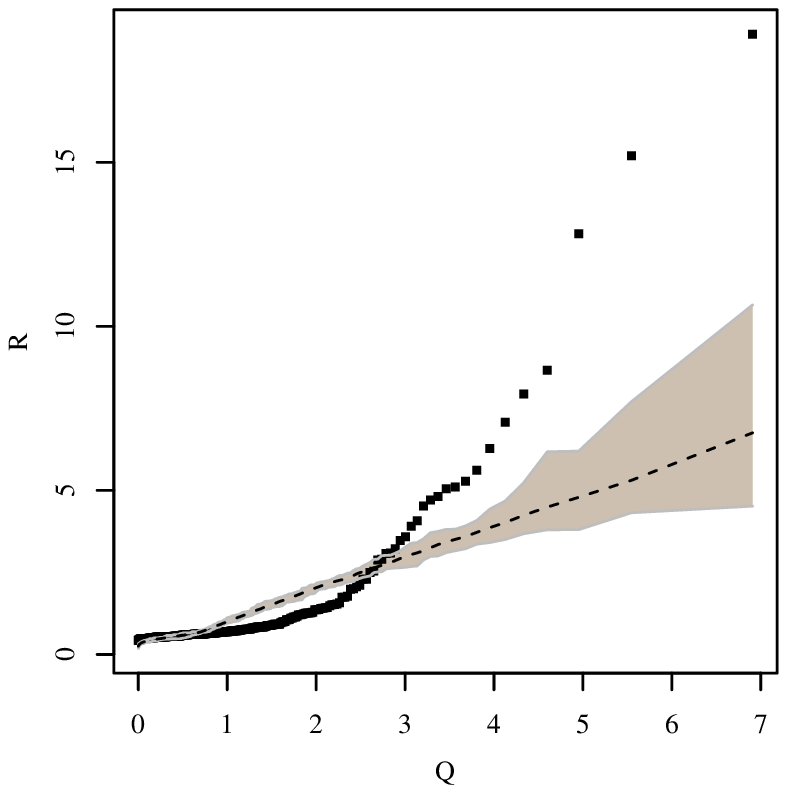}}
\subfigure[tobit-LN]{\includegraphics[height=4cm,width=4cm]{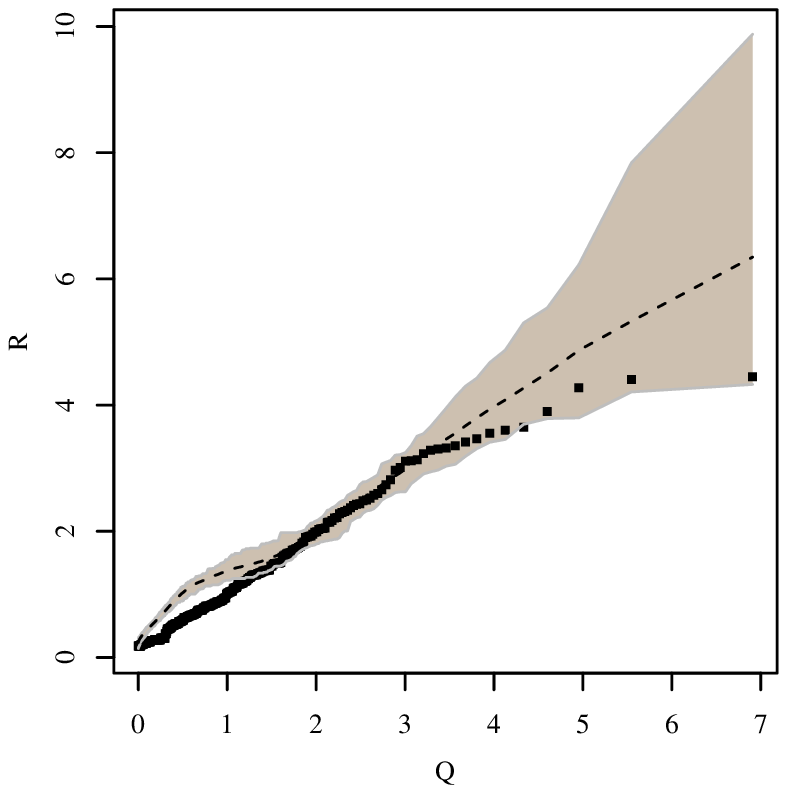}}
\subfigure[tobit-L$t$]{\includegraphics[height=4cm,width=4cm]{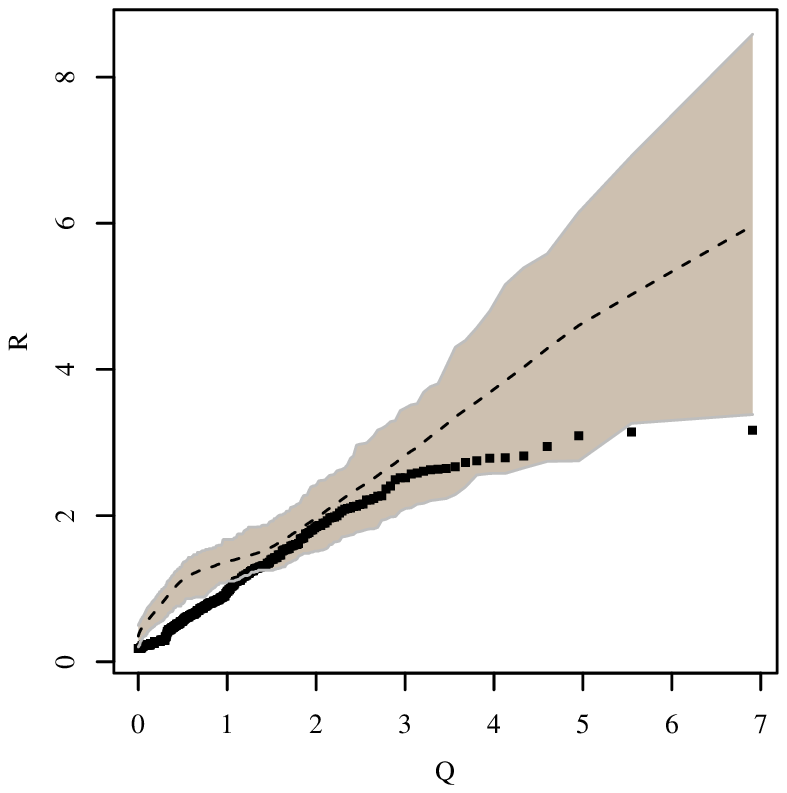}}\\
\subfigure[tobit-LPE]{\includegraphics[height=4cm,width=4cm]{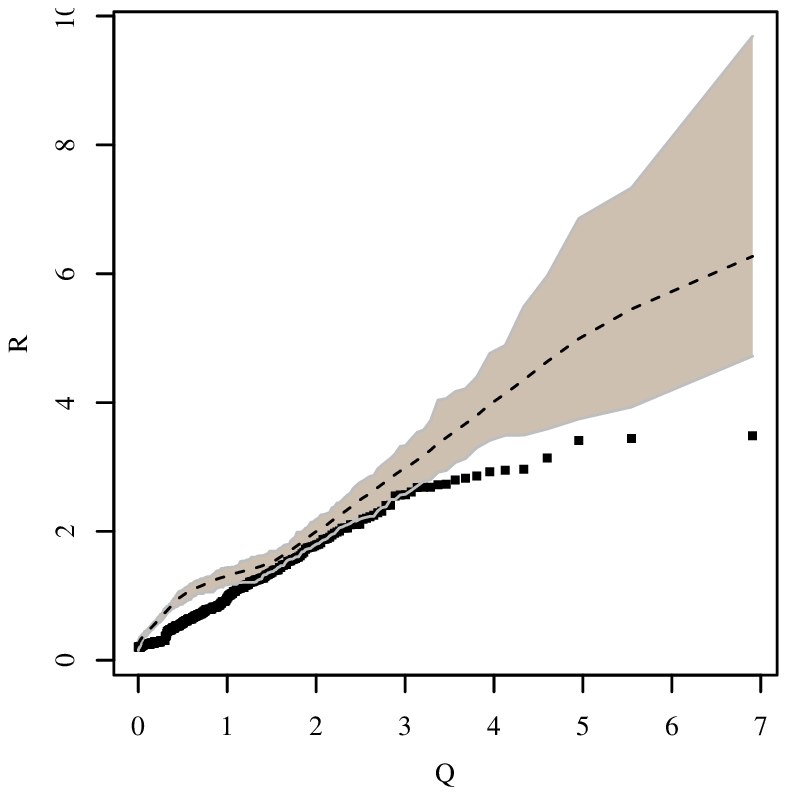}}
\subfigure[tobit-BS]{\includegraphics[height=4cm,width=4cm]{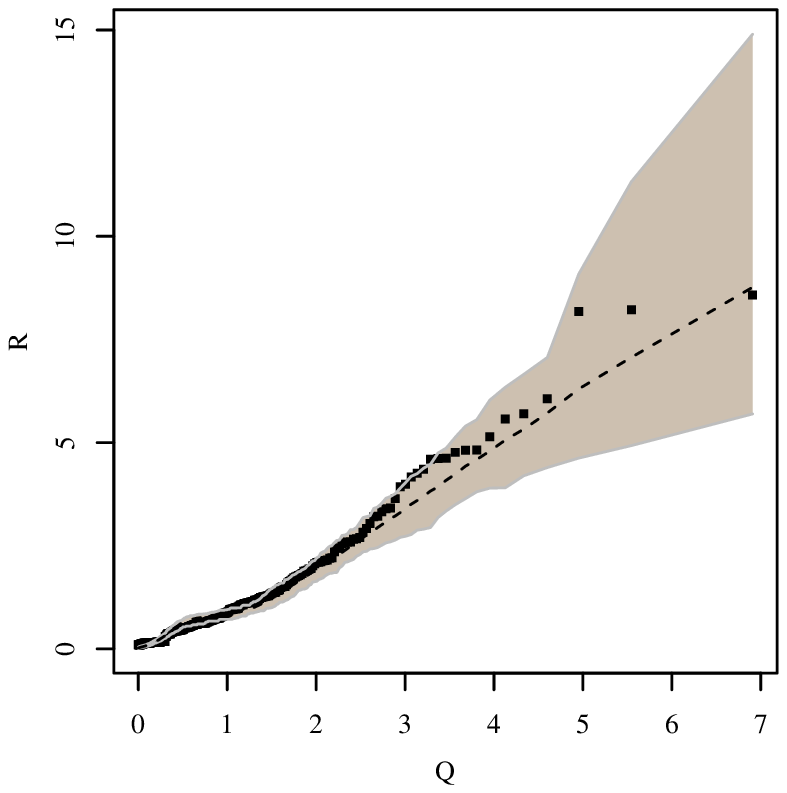}}
\subfigure[tobit-BS-$t$]{\includegraphics[height=4cm,width=4cm]{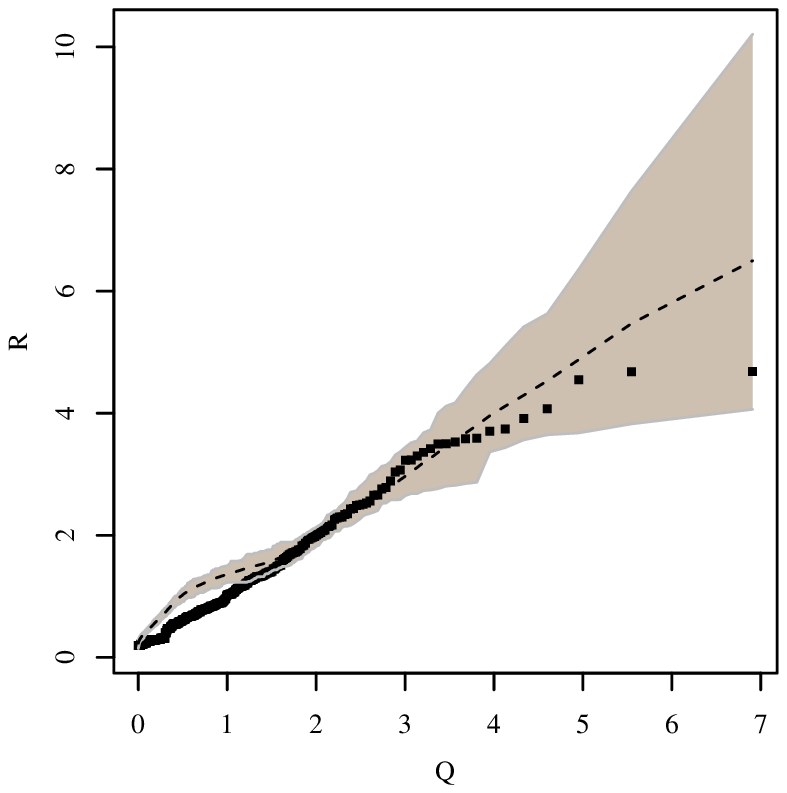}}
\caption{QQ plot and its envelope for the GCS residual for the tobit-log-symmetric models with measles vaccine data.}%
\label{fig:qqplots}
\end{figure}

Figure~\ref{fig:qqplots} displays the quantile versus quantile (QQ) plots with simulated envelope
of the GCS residuals for the tobit-NO, tobit-LN, tobit-L$t$, tobit-LPE, tobit-BS and tobit-BS-$t$ models. This figure indicates 
that GCS residuals in the tobit-log-symmetric models (except the tobit-LPE) show better agreements with the EXP(1) distribution. In special, 
observe a quite good agreement in the tobit-BS case and a poor agreement in the tobit-NO case.

\begin{table}[!ht]
\small
\centering
\caption{ML estimates (with SE in parentheses) and AIC values for the indicated models with the measles vaccine data}
\label{tab:mleresults}
  \begin{tabular}{lcccccccccccccccc} \hline
 Model        & AIC    &   BIC   &&$\phi$        &    $\xi_{1}$    &    $\xi_{2}$    & $\beta_{0}$    & $\beta_{1}$    & $\beta_{2}$     & $\beta_{3}$ \\ \hline
tobit-NO      & 1299.27& 1318.27 &&  0.945       &                &                & 0.597           & 0.225          & $-$0.228        & 0.271 \\[-0.7ex] 
              &        &         &&  (0.047)     &                &                & (0.288)         & (0.297)        & (0.295)         & (0.296) \\[-0.7ex] 
%
%
%
tobit-LN      & 1122.28& 1141.28 &&  1.666       &                &                &  $-$1.239      & 0.315           & 0.138          & 0.087 \\ [-0.7ex]
              &        &         &&  (0.080)     &                &                & (0.184)        & (0.190)         & (0.189)        & (0.189) \\[-0.7ex]
%
%
%
tobit-L$t$    & 1130.68& 1153.47 &&  1.474       &     5          &               &  $-$1.207       & 0.319           & 0.208          & 0.077 \\[-0.7ex]
              &        &         &&  (0.081)     &                &                & (0.183)        & (0.189)         & (0.188)        & (0.189) \\[-0.7ex]
%
%
%
tobit-LPE     & 1123.67&1146.47  &&  1.311       &    0.3         &                &  $-$1.182      & 0.260           & 0.178          & 0.070 \\ [-0.7ex]
              &        &         &&  (0.070)     &                &                & (0.173)        & (0.180)         & (0.175)        & (0.181) \\ [-0.7ex]
%
%
tobit-BS      & 1168.38&1187.37  &&              &    1.545       &                &  $-$0.910      & 0.178           & 0.073          & 0.121 \\ [-0.7ex] 
              &        &         &&              &    (0.081)     &                & (0.105)        & (0.127)         & (0.126)        & (0.126) \\[-0.7ex] 
%
%
tobit-BS-$t$  & 1126.16&1148.96  &&              &    1.662       &    4           &  $-$1.241      & 0.305           & 0.086          & 0.113 \\ [-0.7ex] 
              &        &         &&              &    (0.102)     &                & (0.186)        & (0.191)         & (0.190)        & (0.190) \\ [-0.7ex] 
  \hline
\end{tabular}
\end{table}

\pagebreak

Next, we test the null hypotheses a) ${\cal{H}}_{0}:{\beta}_{1}=0$, 
b) ${\cal{H}}_{0}:{\beta}_{2}=0$ and c) ${\cal{H}}_{0}:{\beta}_{3}=0$, using the LR and {GR} tests. 
We consider only the tobit-LN model as it 
has presented the lowest AIC and BIC values. The corresponding LR and {GR} tests $p$-values are:
a) 0.0970 (LR) and 0.0975 ({GR});
b) 0.4656 (LR) and 0.4657 ({GR});
c) 0.6446 (LR) and 0.6448 ({GR}).

\section{Concluding remarks}\label{sec:6}
We have proposed and analyzed a new class of tobit models for left-censored data. We have considered a likelihood-based 
approach for parameter estimation. We have addressed the issue of performing testing inference in the proposed 
class of tobit models by using the likelihood ratio and gradient statistics. Monte Carlo simulations 
studies were carried out to evaluate the behaviour of the maximum likelihood estimators 
and the likelihood ratio and gradient tests. We have applied the proposed models to a real-world 
data set of measles vaccine data in Haiti. The application has favored 
the use of tobit-log-symmetric models over the classical tobit-normal model.

%
%

\end{document}